\documentstyle[aps,epsfig,multicol,fancyheadings]{revtex}
\begin{document}

\title{Granular fingers in Hele-Shaw experiments}

\author{M.S. Couto, M.L. Martins \& S.G. Alves}

\address{Departamento de F\'{\i}sica, Universidade Federal de Vi\c{c}osa, \\ 36571-000, Vi\c{c}osa, MG, Brazil}

\date{\today}
\maketitle

\vspace{1cm}
\noindent{{\em PACS numbers}: 47.54.+r, 45.70.-n, 45.70.Mg, 45.70.Qj} \\
\noindent{{\em Key-words}: granular materials, Hele-Shaw, pattern formation}
\vspace{1cm}

\begin{multicols}{2}[]
\narrowtext

Granular materials constitute an intermediate state of matter between 
fluids and solids\cite{Jager}. Over the past decade, a large series of
experimental assays on granular systems, concentrated in the quantitative 
studies of compaction, mixing, segregation and patterns formed by 
vibration, have been done [2-9]. They uncovered numerous
 mechanisms (percolation, convection, ordered settling, arching, etc.) 
for segregation of dissimilar grains; remarkable standing-wave patterns 
(stripes, hexagons, disordered waves, cluster of localized solitary 
excitations, etc.) in vibrating layers of granular materials and 
highly inhomogeneous and localized stresses chains in nearly static 
granular media. Nevertheless, in contrast to viscous fingering in
fluids, a widely studied pattern-forming phenomena\cite{Meakin}, 
grain-grain displacement seems to be an unexplored area in granular 
physics. Here we investigate the pattern formation when a grain is 
displaced by another type of grain in a radial Hele-Shaw cell. We 
show that several morphologies can occur, ranging from rounded to 
fingered patterns, interconnected by a continuous crossover. Fourier 
analyses shows that, in contrast to the rounded patterns, the 
fingered shapes present mode selection.

The experiments were carried out in a radial Hele-Shaw cell made of two 
parallel square glass plates, each with a thickness of $0.5$ cm and lengths 
of $90$ cm (bottom plate) and $80$ cm (top plate). The top plate had a hole 
of $0.5$ cm in diameter in its centre for the injection of grains. Initially, 
a dense uniform circular monolayer of styrofoam spheres was formed on the 
cell's bottom plate, always covering the same area. After that, the cell was 
closed with the top plate, over which weights were placed to prevent upward 
movement, and another kind of grain was injected through its hole. In all 
experiments steel spheres with a diameter of $3.94\;\pm\;0.03$ mm were
manually injected, whereas three types of styrofoam initial monolayers were 
used: type $A$, with an average diameter of $2.3\;\pm\;0.1$ mm and low
polydispersion; 
type $B$, composed of spheres with diameters in the range of 0.5 to 4.0 mm and average
diameter of $2.9\;\pm\;0.6$ mm; and type $C$, with an average diameter of
$3.5\;\pm\;0.3$ mm and low polydispersion. After every $100$ g of injected 
spheres, corresponding to the mass of $N\;=\; 395$ spheres, the growing patterns 
were photographed in order to follow their evolution.

Different displacement patterns were observed as a function of the 
diameters of the styrofoam spheres and the plates separation. In Fig. 1, a 
pattern $\sim 50$ cm wide showing well developed fingers, formed after the
injection of $6715$ steel spheres can be seen. In addition to the expected 
interface between the steel/styrofoam spheres, there is an additional 
interface in the displaced grains. This second interface is the border
 between two regions: an outer one, still a dense monolayer, and an inner 
one in which the spheres have moved upward and, hence, is no longer a 
monolayer. All the displacement patterns generated in the experiments 
exhibited this second interface. Also present is a similar upward movement 
of the displacing steel spheres.

In Fig. 2, a qualitative morphological diagram is proposed based on the
data available from the limited range of grains diameters and plates 
separations $D$ tested in our experiments. For large $D$ ($>\;5.60$ mm)
the displacement patterns are circular. For intermediate $D$ (ranging 
from ($5.17$ to $5.60$ mm), the interface of injected steel spheres 
exhibit small wavy or cusp instabilities for type $A$ (small) or type $B$
(intermediate) displaced styrofoam spheres, respectively. For small 
$D$ ($<\;5.17$ mm) fingered patterns are observed, including fingers 
having pronounced tip splitting, resembling the branched erosion 
channels formed in water/glycerine saturated Fointanebleau sand\cite{Mills},
for the smallest $D$. For a given value of $D$, the fingers become 
more apparent as the average diameter of the displaced styrofoam 
spheres is increased. All, these different morphological phases are 
probably not separated by sharp phase transitions and, instead, 
continuous crossovers among them might be observed. Indeed, in Fig. 3, 
the patterns formed by the steel spheres in styrofoam spheres of type 
$B$, suggest a continuous crossover from a fingered pattern (Fig. 3a)
to an almost circular one (Fig. 3d), as the plates separation $D$
increases. The number of injected spheres ($N\;=\;6320$) and the scale
of the drawings are the same for all patterns showed in Fig. 3. From 
now on, the sequence of patterns observed in Fig. 3, which represents 
a cut of constant average diameter (type $B$) in the morphological
diagram will be analysed. Similar behaviours are found for other parallel 
cuts.

From Fig. 4, in which the distance of each point along the border to the 
mass centre of the pattern is represented, one can see the time evolution of 
the patterns depicted in Fig. 3a,d. For the fingered pattern, the formation 
and splitting of the fingers can be clearly seen (Fig. 4a-c). On the contrary, 
for the rounded pattern, although its average radius increases in time, the 
deviation around this value remains small, and fingers are not formed (Fig. 4d-f). 
Figure 5 shows the Fourier transforms of the displacement borders 
in Fig. 4. For the fingered pattern of Fig. 3a, it is possible to see that, as its size 
increases and the fingers become more pronounced, modes are selected, which shift from 
high to low wave numbers (Fig. 5a-c). Rauseo et al.\cite{Rauseo} have observed the same 
behaviour for the power spectra of the curvature versus arc length along the interface 
of the patterns generated by fluid-fluid displacement for their regime of fast flows. 
Conversely, for the pattern of Fig. 3d, which remains rounded during the growth, no mode 
selection is observed (Fig. 5d-f).

In fluid-fluid displacement the fingers are formed when a low viscosity fluid
is injected in a high viscosity one. For the systems studied here, we suspect that the 
role of viscosity is played by the friction of the grains against themselves and the 
plates of the cell. The great elasticity of the styrofoam spheres enhances this friction 
since they can deform and increase their contact area, providing a higher ``effective
viscosity'' as compared to the one for the steel spheres. This increase in friction
becomes more critical as the plates separation is decreased, leading to more pronounced 
fingers in the formed patterns. In fact, experiments in which the styrofoam spheres 
were replaced by hard plastic beads generated less conspicuous fingers as compared 
to the ones formed in styrofoam for the same plates separation.
The relationship between friction and ``effective viscosity'' can explain why finger
instability occurs only after a certain number of steel spheres have been injected. 
At the beginning, the resistance against the displacement front is small since the 
styrofoam spheres, apart from some isolated defects, initially form a monolayer. 
Consequently, its ``effective viscosity'' is low (comparable to that for the steel spheres),
and the border of the displacement pattern is circular with small fluctuations superimposed. 
As the injection proceeds, the styrofoam spheres pile up and can jam in a crescent number of 
points, increasing their ``effective viscosity''. When this value overcomes the ``viscosity''
of the driving grains, the circular interface becomes unstable and fingers begin to form. 
Since the stress generated by the injected grains concentrates at the finger tips, the 
jammed styrofoam spheres in these regions are eventually released due to their elasticity, 
leading to a fast growth of the fingers. In contrast, if hard plastic beads, which can 
be deformed only slightly, are used, the stress concentration at the fingers is not 
sufficient to loose the jammed beads around them, consequently, slowing down the tip 
growth and contributing to stabilise a circular displacement front.

The width of the multilayered region depends on the plates separation. For large 
separations the styrofoam spheres can easily move upward, releasing the stress caused 
by the displacing grain in a narrow region. On the contrary, for small separations the
upward movement of the styrofoam is hindered and a wider region is necessary to release 
the stress. As the pattern develops, the driving front faces an increasing resistance, 
directly observed by the crescent force necessary to inject the steel spheres in the cell. 
At later stages of the experiment, even the displacing grains are under a so large stress 
that the steel spheres themselves begin to move upward, initially in a region close to 
the injection point and them spreading across the pattern. Differently than for the 
styrofoam spheres, even for the largest plates separation used, the steel spheres cannot 
rise enough to form a bilayer. The multilayered region in the styrofoam, as well as in 
the steel spheres, never reaches the outermost ones. There is always a monolayer of 
spheres that seems to be free of stress.

	The main result of this experimental study was the discovery that several 
morphologies, connected by a continuous crossover, can be found for grain-grain displacement 
in a radial Hele-Shaw cell. It was possible to observe that the formation of fingered 
patterns proceeds through the selection and amplification of characteristic wavelengths. 
We are currently refining the morphological diagram and investigating the complex stress 
distribution in this granular system. In particular, the effects of elasticity will be 
the focus of the next experiments.

M.L. Martins is funded by CNPq, Brazilian Agency.

Correspondence and requests for materials should be addressed to M.S. Couto (e-mail: 
mscouto@mail.ufv.br).

\end{multicols}
\widetext

\begin{figure}
\centerline{\epsfig{file=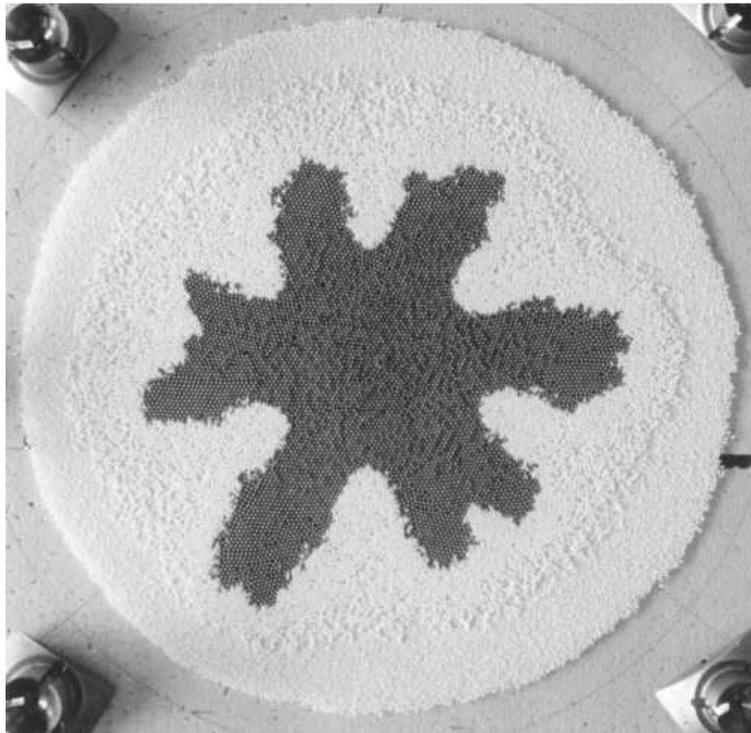,width=10cm,angle=-0}}
\caption{Fingered pattern formed by the injection of steel spheres in a monolayer of
styrofoam spheres of type $B$. The plates separation is $4.93\;\pm\;0.04$ mm.}
\label{fig1}
\end{figure}

\begin{figure}
\centerline{\epsfig{file=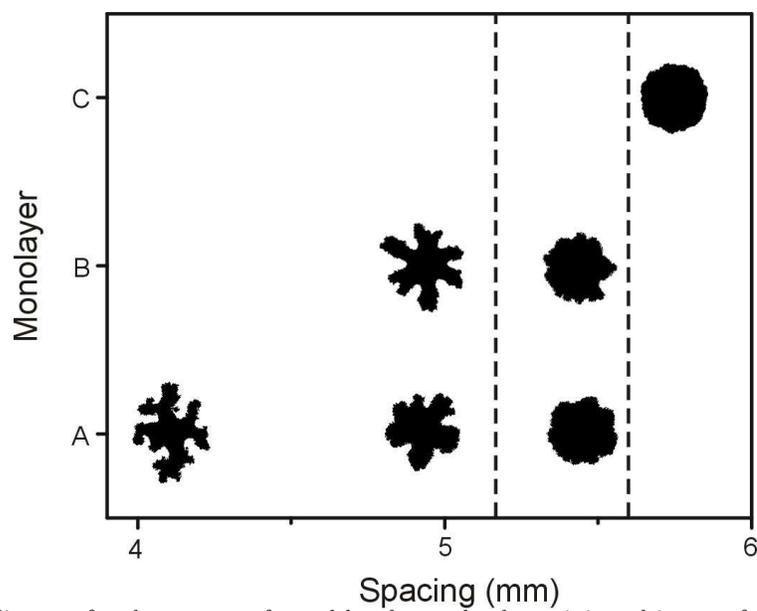,width=10cm,angle=-0}}
\caption{Morphological diagram for the patterns formed by the steel spheres injected in
styrofoam particles as a function of the diameter of the styrofoam spheres and the plates separation.}
\label{fig2}
\end{figure}

\begin{figure}
\centerline{\epsfig{file=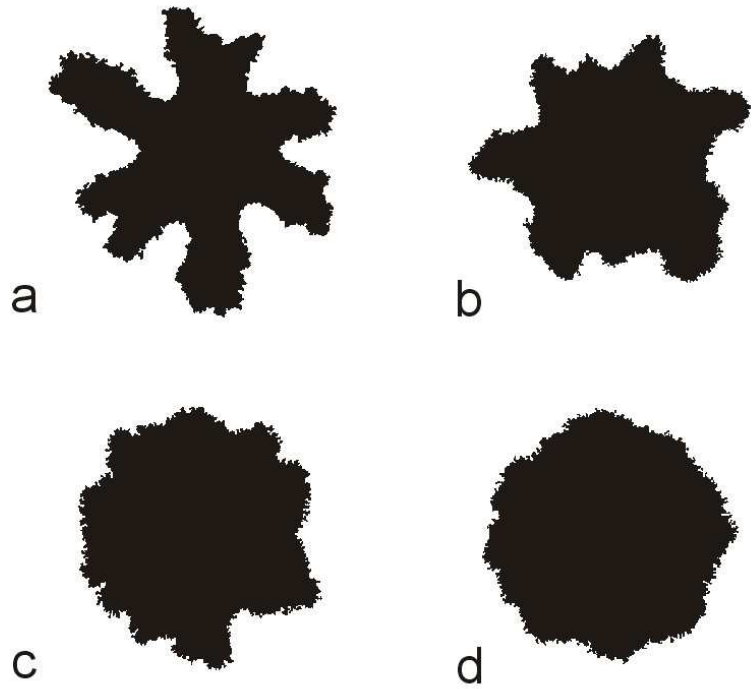,width=10cm,angle=-0}}
\caption{Morphological patterns formed by the steel spheres. The plates separation $D$
 are $4.93$ mm (a), $5.17$ mm (b), $5.44$ mm (c) and $5.73$ mm (d). The initial monolayer is
of type $B$.}
\label{fig3}
\end{figure}

\begin{figure}
\centerline{\epsfig{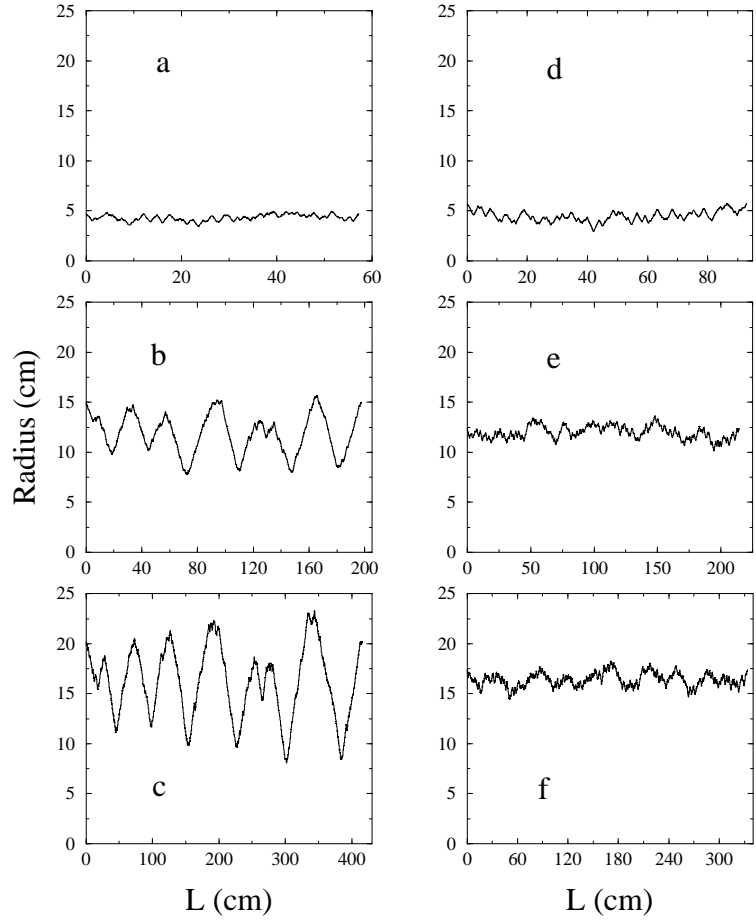}}
\caption{The border of the patterns shown in Fig. 3a ({\bf a-c}) and Fig. 3d ({\bf d-f}) for
three different numbers of injected grains. {\bf a,d}, $N\;=\;395$. {\bf b,e}, $N\;=\;3160$. {\bf c,f}, $N\;=\;5925$.}
\label{fig4}
\end{figure}

\begin{figure}
\vspace{5cm}
\centerline{\epsfig{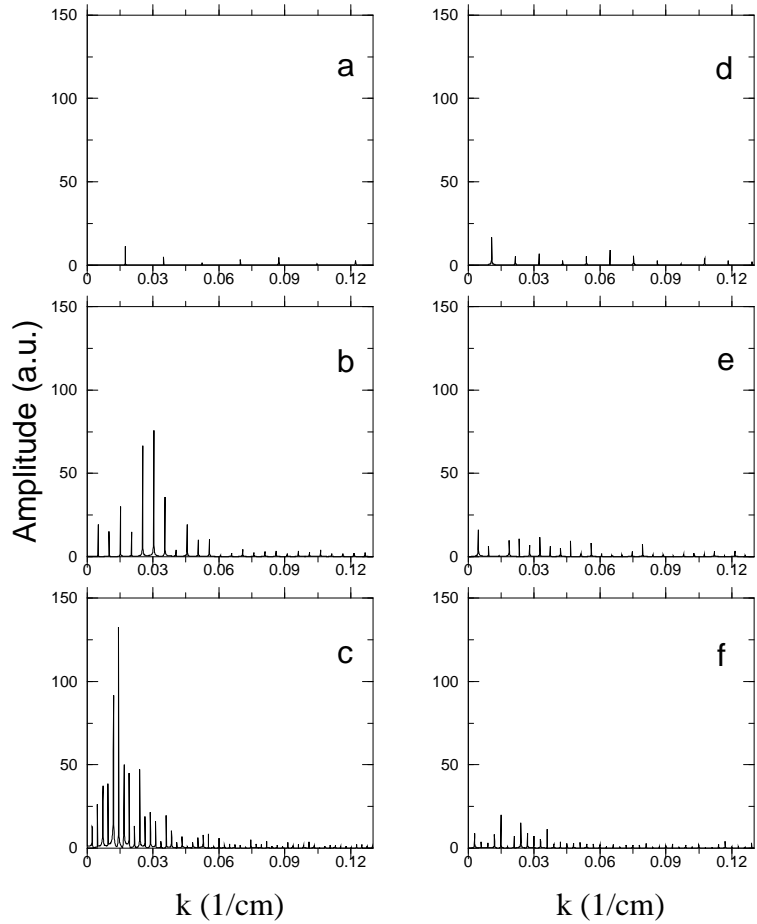}}
\caption{Power spectra of the border of the patterns shown in Fig. 4. {\bf a-c}, Transforms for
Fig. 4a-c. The two strongest modes selected correspond to the formation of fingers and 
their tip splitting. {\bf d-f}, Transforms for Fig. 4d-f.}
\label{fig5}
\end{figure}

\end{document}